\documentclass[preprint,12pt]{elsarticle}
\usepackage{graphicx}
\usepackage{subfigure}
\usepackage{amssymb}
\journal{Solid State Communications}

\begin{document}
\begin{frontmatter}

\title{
Chemical Trend of Superconducting Transition Temperature
 in Hole-doped CuBO$_2$, CuAlO$_2$, CuGaO$_2$ and CuInO$_2$ 
}

\author{Akitaka Nakanishi\corref{cor}}
\ead{nakanishi@aquarius.mp.es.osaka-u.ac.jp}
\author{Hiroshi Katayama-Yoshida}
\cortext[cor]{Tel.: +81-6-6850-6504. Fax: +81-6-6850-6407.}
\address{Department of Materials of Engineering Science,
 Osaka University, Toyonaka, Osaka 560-8531, Japan}

\begin{abstract}
We calculated the superconducting transition temperature $T_{\rm c}$ of
 hole-doped CuBO$_2$, CuAlO$_2$, CuGaO$_2$ and CuInO$_2$ using first-principles.
The calculated $T_{\rm c}$ are about 50 K for CuAlO$_2$, 10 K for CuBO$_2$ and CuGaO$_2$
 and 1 K for CuInO$_2$ at maximum in the optimum hole-doping concentration.
The low $T_{\rm c}$ of CuInO$_2$ is attributed to
 the weak electron-phonon interaction caused by the low covalency
 and heavy atomic mass.
\end{abstract}


\end{frontmatter}

\section{Introduction}
CuAlO$_2$ has a delafossite structure (Left of Fig. \ref{fig:str})
 and a two-dimensional electronic structure
 caused by the natural super-lattices of O-Cu-O dumbbell.
Kawazoe {\it et al.} have found that
 the CuAlO$_2$ is $p$-type transparent
 conducting oxides (TCO) without any intentional doping. \cite{Kawazoe1997}
Nakanishi {\it et al.} studied the pressure dependence of the structures
 \cite{Nakanishi2011a} and the role of the self-interaction correction
 in CuAlO$_2$. \cite{Nakanishi2011b}
Transparent $p$-type conductors such as CuAlO$_2$ are
 important for the $p$-$n$ junction of TCO
 and a realization of high-efficiency photovoltaic solar-cells.
First-principles calculations have shown the possibility
 for high efficiency thermoelectric power application with about 1\% hole-doping.
 \cite{Funashima2004,Yoshida2009,Hamada2006}

Katayama-Yoshida {\it et al.} have simulated
 the Fermi surface of the hole-doped CuAlO$_2$
 by shifting the Fermi level rigidly
 and proposed that the nesting Fermi surface may cause
 a strong electron-phonon interaction
 thus a transparent superconductivity for visible light. \cite{Yoshida2003}
However,
 they have not calculated the superconducting transition temperature $T_{\rm c}$.
In previous study,
 we calculated the $T_{\rm c}$ of hole-doped CuAlO$_2$ \cite{Nakanishi2012a}
 and found that the $T_{\rm c}$ increases up to about 50\,K
 due to the strong electron-phonon interaction
 by the two dimensional flat valence band.
The origin of the flat band is the $\pi$-band of
 hybridized O 2p$_z$ and Cu 3d$_{3z^2-r^2}$
 on the frustrated triangular lattice in the two dimensional plane.

It is interesting to see the relation between the $T_{\rm c}$
 and the flatness of the flat band by changing the Cu 3d$_{3z^2-r^2}$,
 Ag 4d$_{3z^2-r^2}$ and Au 5d$_{3z^2-r^2}$.
In the next study,
 we calculated the $T_{\rm c}$ of hole-doped delafossite AgAlO$_2$ and AuAlO$_2$. \cite{Nakanishi2012b}
The calculated $T_{\rm c}$ are about 40 K for AgAlO$_2$ and 3 K for AuAlO$_2$
 at maximum in the optimum hole-doping concentration.
The low $T_{\rm c}$ of AuAlO$_2$ is attributed to
 the weak electron-phonon interaction caused by the low covalency
 and heavy atomic mass.
In this study,
 we calculated $T_{\rm c}$ and the electron-phonon interaction
 versus the chemical trend of hole-doped CuBO$_2$, CuAlO$_2$, CuGaO$_2$ and CuInO$_2$.

\section{Calculation Methods}
The calculations were performed
 within the density functional theory \cite{Hohenberg1964,Kohn1965}
 with a plane-wave pseudopotential method,
 as implemented in the Quantum-ESPRESSO code. \cite{Giannozzi2009}
We employed the Perdew-Wang 91 \cite{Perdew1992} for CuBO$_2$, CuAlO$_2$ and CuGaO$_2$
 and the Perdew-Burke-Ernzerhof \cite{Perdew1996} for CuInO$_2$
 generalized gradient approximation (GGA) exchange-correlation functional
 and ultra-soft pseudopotentials. \cite{Vanderbilt1990}
For the pseudopotentials,
 $d$ electrons of transition metals were also included
 in the valence electrons.
In reciprocal lattice space integral calculation,
 we used $8\times8\times8$ (electron and phonon)
 and $32\times32\times32$ (density of states and average at Fermi level)
 ${\bf k}$-point grids in the Monkhorst-Pack grid. \cite{Monkhorst1976}
The energy cut-off for wave function was 40  Ry and
 that for charge density was 320  Ry.
These ${\bf k}$-point grids and cut-off energies are fine
 enough to achieve convergence within 10 mRy/atom in the total energy.

\begin{figure}[htbp]
  \begin{center}
    \includegraphics{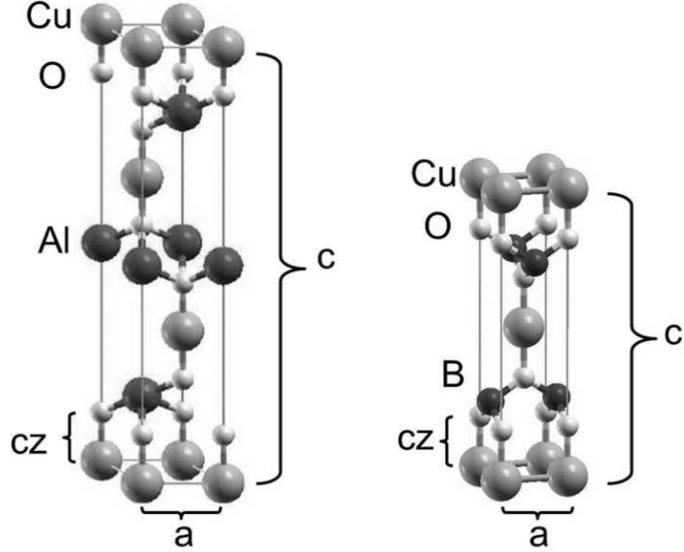}
    \caption{The crystal structure of delafossite CuAlO$_2$ (left)
         and body-centered tetragonal CuBO$_2$ (right).}
    \label{fig:str}
  \end{center}
\end{figure}

The delafossite structure belongs to the space group R$\bar3$m (No. 166)
 and is represented by cell parameters $a$ and $c$,
 and internal parameter $z$ (Left of Fig. \ref{fig:str}).
These parameters were optimized by 
 the constant-pressure variable-cell relaxation using
 the Parrinello-Rahman method \cite{Parrinello1980}
 without any symmetry constraints.

As it is difficult for first-principles to exactly deal with doped systems.
 we had to implement some approximations.
Let us take the electron-phonon interaction $\lambda$ for example.
$\lambda$ is defined as follows:

\begin{equation}
 \lambda
 =
 \sum_{\nu{\bf q}}
 \frac{
 2N(\varepsilon_{\rm F})
 \sum_{{\bf k}}|M_{{\bf k,k+q}}^{\nu{\bf q}}|^2
 \delta(\varepsilon_{\bf k}-\varepsilon_{\rm F})
 \delta(\varepsilon_{{\bf k+q}}-\varepsilon_{\rm F})
 }
 {
 \omega_{\nu{\bf q}}\sum_{{\bf kq'}}
 \delta(\varepsilon_{\bf k}-\varepsilon_{\rm F})
 \delta(\varepsilon_{{\bf k+q'}}-\varepsilon_{\rm F})
 }.
\label{eqn:lambda}
\end{equation}
(1)
 For the non-doped systems,
 we calculated the dynamical matrix,
 the phonon frequency $\omega_{\nu{\bf q}}$ 
 and the electron-phonon matrix $M_{{\bf k,k+q}}^{\nu{\bf q}}$.
(2)
 For the doped systems,
 we calculated the Fermi level $\varepsilon_{\rm F}$ and
 the density of states at the Fermi level $N(\varepsilon_{\rm F})$
 with the number of valence electrons reduced
 using the eigenvalues $\varepsilon_{\bf k}$ of the non-doped system.
(3)
 By using the results of (1) and (2),
 we calculated the electron-phonon interaction $\lambda$
 and the other superconducting properties.
This approximation is based on
 the idea that the doping does not greatly change
 electron and phonon band structures.
In this study,
 we show the results of $0.1\sim1.0$ hole-doped systems.

We calculated the superconducting transition temperature
 by using the Allen-Dynes modified McMillan formula. \cite{McMillan1968,Allen1975} 
According to this formula,
 $T_{\rm c}$ is given by three parameters:
 the electron-phonon interaction $\lambda$, 
 the logarithmic averaged phonon frequency $\omega_{\log}$, 
 and the screened Coulomb interaction $\mu^{\ast}$, in the following form.
\begin{eqnarray}
  T_{\rm c}&=&\frac{\omega_{\log}}{1.2}
  \exp \left( \frac{-1.04(1+\lambda )}
  {\lambda-\mu^{\ast}(1+0.62\lambda )} \right). \\
  \omega_{\log} 
  &=&\exp \left(\frac{2}{\lambda}\int_0^\infty
  d\omega\frac{\alpha^2F(\omega)}{\omega}\log\omega\right).
\end{eqnarray}
Here, $\alpha^2F(\omega)$ is the Eliashberg function.
$\lambda$ and $\omega_{\rm log}$ are obtained by 
 the first-principle calculations using
 the density functional perturbation theory. \cite{Baroni2001}
As for $\mu^{\ast}$,
 we assume the value $\mu^{\ast}=0.1$.
This value holds for weakly correlated materials.

\section{Calculation Results and Discussion}
First,
 we optimized the cell parameters.
The results show that
 the optimized structure is delafossite
 as no structural transition occured.
However,
 the phonon frequency of CuBO$_2$ is negative.
It means that the structure is only locally stable.
Then after moving some atoms from their initial positions and optimizing the structure again,
 we found out that CuBO$_2$ relaxed to a body-centered tetragonal structure.
This structure is represented by cell parameters $a$ and $c$,
 and internal parameter $z$ (Right of Fig. \ref{fig:str}).
Table \ref{tab:opt}
 shows the optimized cell parameters.

\begin{table} [htpd]
\begin{center}
  \begin{tabular}{ccccc}
  \hline
            & CuBO$_2$ & CuAlO$_2$ & CuGaO$_2$ & CuInO$_2$ \\ \hline 
  $a$ [\AA] & 2.534    & 2.859     & 3.002     & 3.367     \\
  $c/a$     & 4.253    & 5.965     & 5.759     & 5.250     \\
  $z$       & 0.174    & 0.110     & 0.108     & 0.106     \\
  \hline
  \end{tabular}
  \caption{The optimized cell parameters of
           body-centered tetragonal CuBO$_2$
           and delafossite CuAlO$_2$, CuGaO$_2$ and CuInO$_2$.}
  \label{tab:opt}
\end{center}
\end{table}

\begin{figure}[htbp]
\begin{center}
  \includegraphics{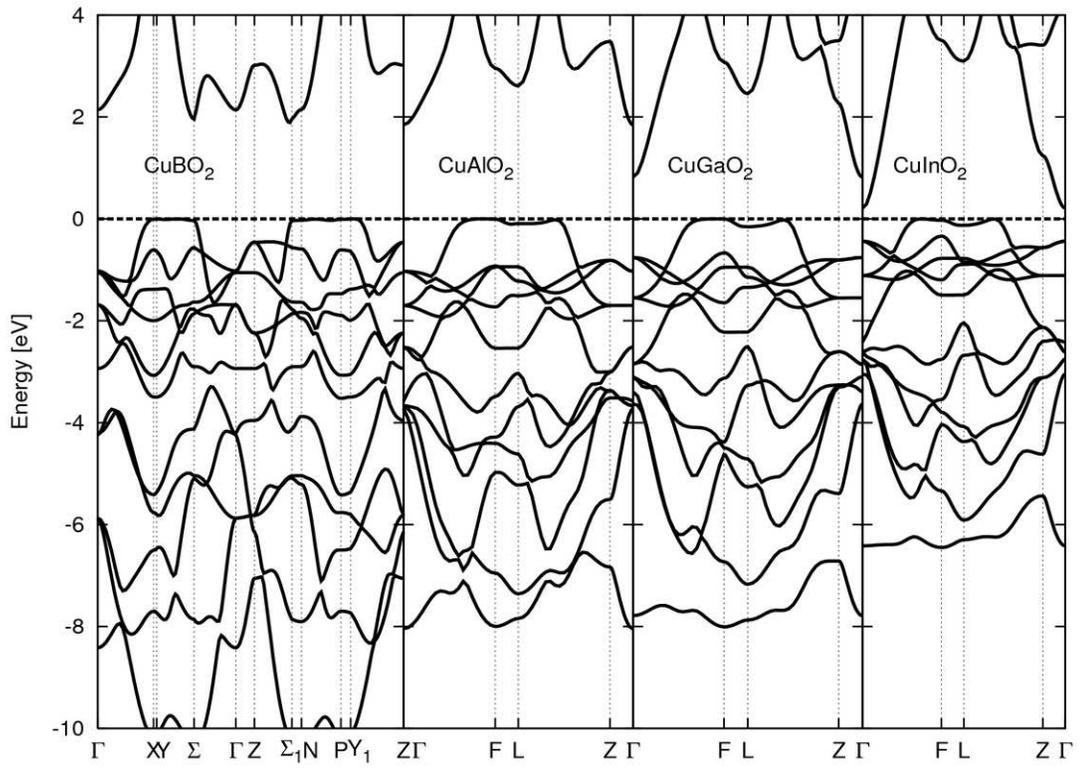}
  \caption{Band structures of CuBO$_2$, CuAlO$_2$, CuGaO$_2$ and CuInO$_2$.}
  \label{fig:band}
\end{center}
\end{figure}

\begin{figure}[htbp]
\begin{center}
  \includegraphics{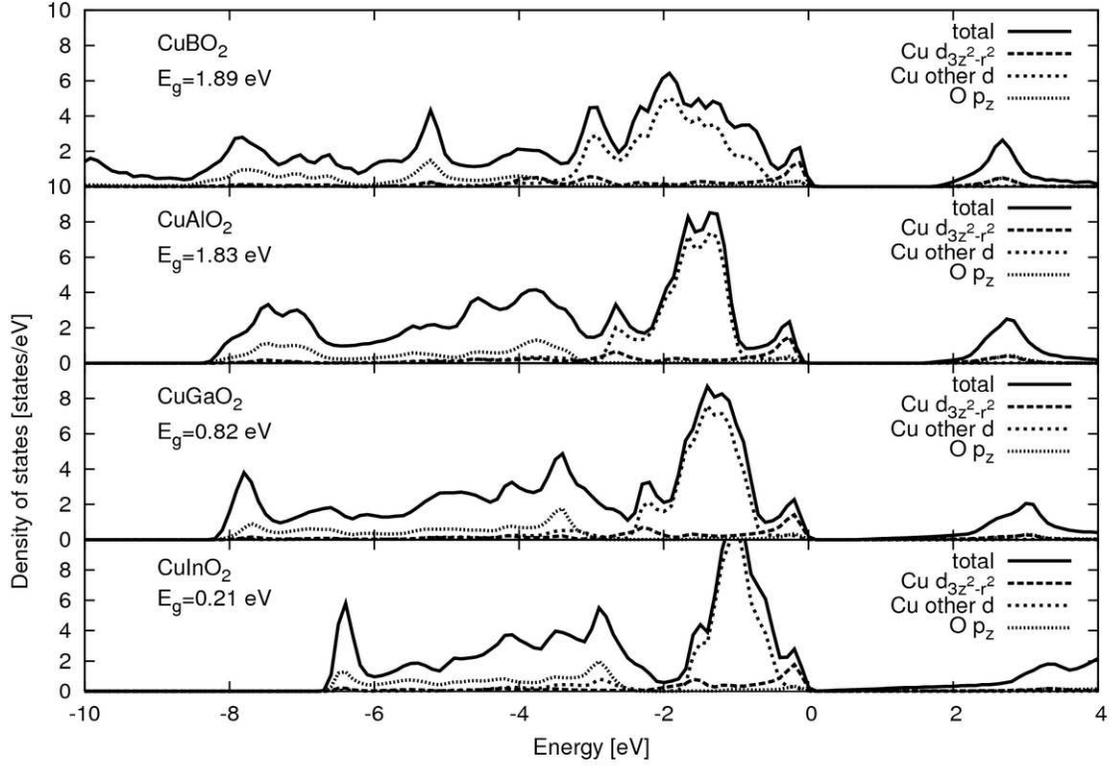}
  \caption{Density of states and energy gaps.}
  \label{fig:dos}
\end{center}
\end{figure}

Figures \ref{fig:band} and \ref{fig:dos} show
 the band structures, the densities of states (DOS) and energy gaps.
The energy gaps are 1.89 eV for CuBO$_2$, 1.83 for CuAlO$_2$, 0.82 for CuGaO$_2$ and 0.21 for CuInO$_2$. 
Small gaps of CuGaO$_2$ and CuInO$_2$ show that these covalent bonding is weak.
The target materials have the flat valence bands and small peaks of DOS
 due to the two dimensionality in O-Cu-O dumbbell array.
These peaks are mainly constructed
 by the two-dimensional $\pi$-band of 
 Cu 3$d_{3z^2-r^2}$-O 2$p_z$ anti-bonding states.
CuBO$_2$ has wider non-covalent Cu $d$-band and a narrower peak than CuAlO$_2$.
This means that the covalency of CuBO$_2$ is lower than that of CuAlO$_2$
 though their energy gaps are the same.

\begin{figure}[htbp]
\begin{center}
  \includegraphics{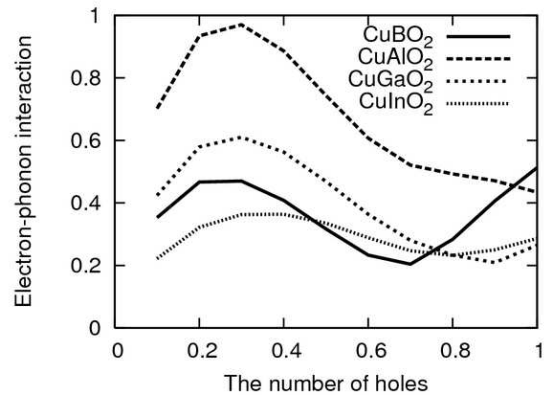}
  \caption{Electron-phonon interaction $\lambda$.}
  \label{fig:lambda}
\end{center}
\end{figure}

Figure \ref{fig:lambda} shows
 the electron-phonon interaction $\lambda$. 
There are peaks at the number of holes $N_{\rm h}=0.3$.
In CuBO$_2$,
 a first peak is narrow and a second peak
 which is due to the narrow DOS peak appears
 in the heavily doped region of $N_{\rm h}=0.7\sim1.0$.
CuAlO$_2$ has higher $\lambda$ than CuBO$_2$, CuGaO$_2$ and CuInO$_2$.
The difference in $\lambda$ strength is due to the covalency mentioned above.

\begin{figure}[htbp]
\begin{center}
  \includegraphics{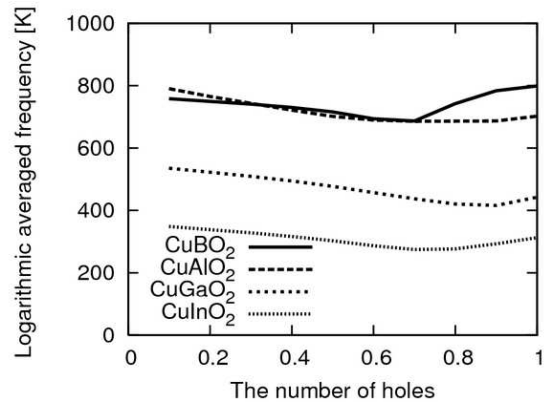}
  \caption{Logarithmic averaged phonon frequency $\omega_{\log}$.}
  \label{fig:omega}
\end{center}
\end{figure}

Figure \ref{fig:omega} shows
 the logarithmic averaged phonon frequency $\omega_{\log}$.
These are almost constant for the number of holes.
The difference of $\omega_{\log}$ is
 mainly due to atomic mass.
Therefore,
 CuGaO$_2$ and CuInO$_2$ have smaller $\omega_{\log}$
 than CuBO$_2$ and CuAlO$_2$.
In the lightly doped region,
 CuBO$_2$ and CuAlO$_2$ have almost the same $\omega_{\log}$
 though Al is heavier than B.
This is due to large electron-phonon interaction of CuAlO$_2$.

\begin{figure}[htbp]
\begin{center}
  \includegraphics{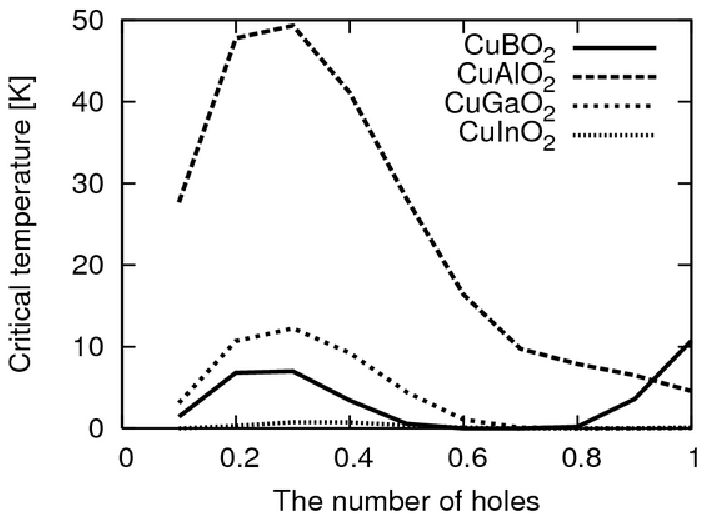}
  \caption{Superconducting transition temperature $T_{\rm c}$.}
  \label{fig:tc}
\end{center}
\end{figure}

Figure \ref{fig:tc} shows
 the superconducting transition temperature $T_{\rm c}$.
Since CuInO$_2$ has very low $T_{\rm c}$ ($<1\,K$),
 its curve is no longer invisible in Fig. \ref{fig:tc}.
The $T_{\rm c}$ variation is determined
 mainly by the electron-phonon interaction
 because logarithmic averaged phonon frequencies $\omega_{\log}$
 are almost constant for the number of holes.
CuBO$_2$, CuGaO$_2$ and CuInO$_2$ have much lower $T_{\rm c}$ than CuAlO$_2$
 because they have low covalency and $\lambda$ as mentioned above.

\section{Conclusions}
In summary,
 we calculated the chemical trend of superconducting transition temperature of
 the hole-doped delafossite CuBO$_2$, CuAlO$_2$, CuGaO$_2$ and CuInO$_2$.
The calculated $T_{\rm c}$ are about 50 K for CuAlO$_2$, 10 K for CuBO$_2$ and CuGaO$_2$
 and 1 K for CuInO$_2$ at maximum in the optimum hole-doping concentration.
The low $T_{\rm c}$ of CuInO$_2$ is attributed to
 the weak electron-phonon interaction caused by the low covalency
 and heavy atomic mass.

\section*{Acknowledgment}
The authors acknowledge the financial support from
 the Global Center of Excellence (COE) program "Core Research and
 Engineering of Advanced Materials - Interdisciplinary Education Center for
 Materials Science", the Ministry of Education, Culture, Sports, Science and
 Technology, Japan, and a Grant-in-Aid for Scientific Research on Innovative
 Areas "Materials Design through Computics: Correlation and Non-Equilibrium Dynamics".
We also thank to the financial support from the Advanced Low Carbon Technology Research and 
Development Program, the Japan Science and Technology Agency for the financial support.

\bibliographystyle{elsarticle-num}
\bibliography{bibfile}

\end{document}